\documentclass[reprint,twocolumn,amsmath,amssymb,aps,prl,floatfix,superscriptaddress]{revtex4-2}

\usepackage{amsmath}
\usepackage{amssymb}

\usepackage{graphicx}               % enables inclusion of figure files
\usepackage{wrapfig}

\usepackage{dcolumn}                % enables aligning table columns on decimal point

\usepackage{upgreek}                % enables Greek in roman font (for micro- prefix), 2022-05-27: added by KL
\usepackage{bm}                     % enables bold math. used for symbols \cal F, \Omega and \ell 

\usepackage{xcolor}
\usepackage{hyperref}               % adds hypertext capabilities

\usepackage{accents}
\usepackage{wasysym}                % for diameter symbol, 2017-08-07: added by KL

\usepackage{lineno}
\usepackage{mathrsfs}
\usepackage{comment}
\usepackage{mathtools}
\usepackage[version=4]{mhchem}
\usepackage{float}
\usepackage{enumitem}
\usepackage{tikz}

\usepackage[normalem]{ulem} % allows strikeout for review editing, 2022-07-20 added by KL

\hypersetup{colorlinks,linkcolor={blue!90!black},citecolor={blue!90!black},urlcolor={blue!90!black}}

\begin{document}

\title{Phototactic Decision-Making by Micro-Algae}

\author{Shantanu Raikwar}
\affiliation{Laboratoire de Physique de l'\'Ecole Normale Sup\'erieure, ENS, Universit\'e PSL, CNRS, Sorbonne Universit\'e, Universit\'e Paris Cit\'e, F-75005 Paris, France}
\author{Adham Al-Kassem}
\affiliation{Laboratoire de Physique de l'\'Ecole Normale Sup\'erieure, ENS, Universit\'e PSL, CNRS, Sorbonne Universit\'e, Universit\'e Paris Cit\'e, F-75005 Paris, France}
\author{Nir S. Gov}
\affiliation{Department of Chemical and Biological Physics, Weizmann Institute of Science, Rehovot 76100, Israel}
\affiliation{Department of Physiology, Development and Neuroscience, University of Cambridge, Cambridge CB2 3DY, United Kingdom}
\author{Adriana I. Pesci}
\affiliation{Department of Applied Mathematics and Theoretical Physics, University of Cambridge, Wilberforce Road, Cambridge CB3 0WA, United Kingdom}
\author{Rapha\"el Jeanneret}
\affiliation{Laboratoire de Physique de l'\'Ecole Normale Sup\'erieure, ENS, Universit\'e PSL, CNRS, Sorbonne Universit\'e, Universit\'e Paris Cit\'e, F-75005 Paris, France}
\author{Raymond E. Goldstein}
\affiliation{Department of Applied Mathematics and Theoretical Physics, University of Cambridge, Wilberforce Road, Cambridge CB3 0WA, United Kingdom}
\email[Correspondence:\\]{raphael.jeanneret@phys.ens.fr  R.E.Goldstein@damtp.cam.ac.uk}

\date{\today}

\begin{abstract}  
We study how simple eukaryotic organisms make decisions 
in response to competing stimuli in the context of phototaxis by the unicellular  
alga {\it Chlamydomonas reinhardtii}.  While negatively 
phototactic cells swim directly away from a 
collimated light beam, 
when presented with two beams of adjustable intersection angle and intensities, we find that cells swim in a direction 
given by an intensity-weighted average of the two light propagation vectors.  This geometrical law is
a fixed point of an adaptive 
model of phototaxis and minimizes the average light intensity falling on the anterior 
pole of the cell. At large angular separations, 
subpopulations of cells swim away from
one source or the other, or along the direction of 
the geometrical law, with some cells stochastically switching between the three directions.  This 
behavior is shown 
to arise from a population-level distribution of 
photoreceptor locations that
breaks front-back symmetry of photoreception. 
\end{abstract}

\maketitle

In areas as diverse as ecology \cite{FretwellLucas1969,Harper1982}, microbiology \cite{Adler1974}, 
evolutionary biology and the psychology of human behavior \cite{SantosRosati2015}
the question arises of how individuals make decisions when confronted with competing environmental stimuli.  For complex organisms with a highly developed neural system, such decision-making 
may involve weighing the costs and benefits of the choices along with a balance between 
immediate rewards and long-term consequences.  
The situation is less clear for aneural organisms such as plants, bacteria and amoebae,
but they appear to utilize similar mechanisms \cite{Reid2015}.

The simplest setting for decision-making clearly involves just two choices.  At the scale of 
microorganisms there have been studies of the chemotactic response of the bacterium 
\textit{E. coli} to two opposing chemical stimuli \cite{Adler1974}, and the phototactic response of 
colonies of cyanobacteria to two light sources \cite{Chau2017,Kim2017,Menon2020}.  These have
suggested a ``summation rule" in which the addition (scalar or vector) of the
two stimuli forms the basis of the decision.  Similar rules have been found in the study of plants 
\cite{Riviere2023}.

Of the many types of taxes exhibited by microorganisms\textemdash chemotaxis, phototaxis, 
viscotaxis, durotaxis\textemdash phototaxis is distinguished by the fact that the 
stimulus direction and magnitude can be changed arbitrarily fast, without the complexities
of a diffusive process or intervening surfaces.  In the case of algal phototaxis, a long history of 
studies \cite{Foster1980,Ruffer1990,Ruffer1991,Schaller1997,Hegemann_vision,Josef2005,Josef2006,Hegemann_review,Jekely2009,Drescher2010,Bennett2015,GoniumPRE,Leptos2023} has shown that the light-sensing process is ``line of sight" in the sense that
each cell has a photosensor that responds when directly illuminated, triggering 
changes in flagellar beating that produce alignment with the light.  The two key ingredients
for accurate phototaxis are the spinning of cells about a body-fixed axis and the directionality of
the photoreceptor, achieved by a protein layer (the ``eyespot") that blocks light coming from behind
the cell.  Flagellar beating exhibits a rapid response to changes in light and a slower adaptation that is tuned to the 
spinning period \cite{,Josef2005,Josef2006,Leptos2023,Yoshimura2001}, and  
cells can exhibit positive or negative phototaxis depending on the prevailing
light intensity.   

\begin{figure*}[t]
\includegraphics[width=2.0\columnwidth]{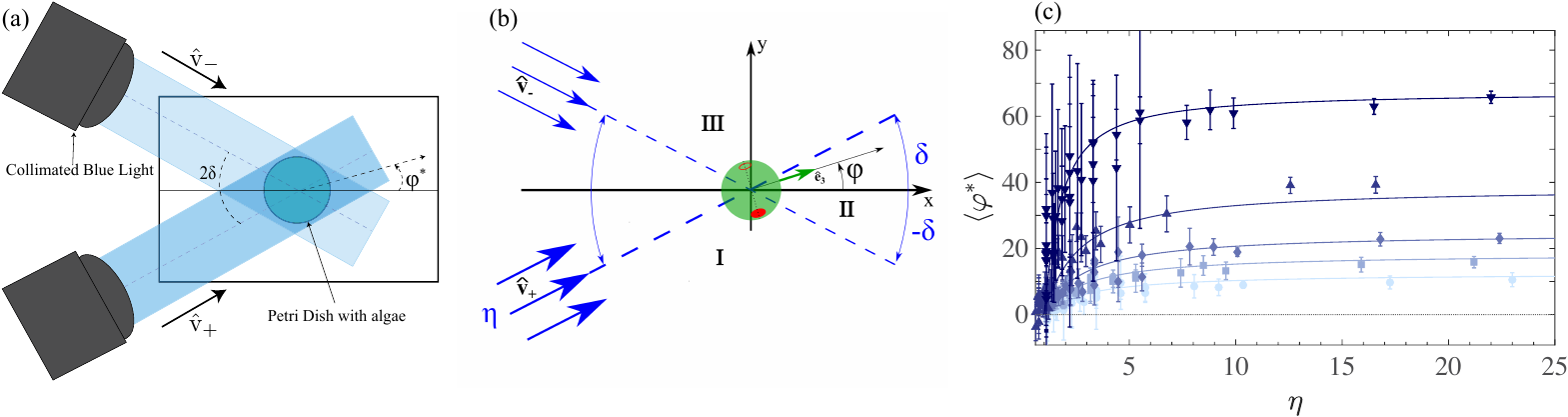}
\caption{Experimental results. (a) Setup: two collimated lights shine toward the $x$-axis at angles $\pm \delta$, the lower with intensity $\eta$ relative to the upper.  (b) 
A swimming cell whose axis $\hat{\bf e}_3$ 
is at an angle $\varphi$ with respect to the $x-$axis. Eyespot at two instants in time separated by a half cycle 
is shown as a solid and open red ellipse. 
(c) Swimming angle as a function 
of intensity ratio $\eta$ for 
$\delta=12.5^\circ$, $18.45^\circ$, 
$24.74^\circ$, $38.4^\circ$, and 
$67.6^\circ$ increasing upwards, along with the theoretical
prediction \eqref{tangentlaw} for each value of $\delta$. }
\label{fig1}
\end{figure*}

Here we report on an extensive investigation at the single cell level of phototactic decision-making 
by unicellular green algae. By employing the experimental setup shown in Fig.~\ref{fig1}(a,b),
in which two collimated light beams with
independently adjustable intensities intersect at a prescribed angle within a dilute suspension of 
negatively phototactic \textit{Chlamydomonas reinhardtii}, we track thousands of individual cells' decisions on the choice
of swimming direction.  The steady-state swimming directions are found to follow what we call the ``tangent law", 
an intensity-weighted average of the two light propagation vectors, and we show that this law is the fixed point of an adaptive 
model of phototaxis \cite{Drescher2010,GoniumPRE,Leptos2023}.  Studies of the
response of cells to rapid changes in light direction reveal a surprisingly fast cell reorientation that
can be quantitatively described as a limit of the adaptive theory.  Finally, motivated by bifurcation phenomena
found in certain decision-making processes \cite{Sridhar2021}, we examine swimming trajectories when 
the two lights are nearly
antiparallel and find that there are three distinct subpopulations of cells: those that either (i) swim away from one 
source or from the other, (ii) go
along the direction of the geometrical law, 
and (iii) exhibit stochastic switching between the first two choices.  We show that this behavior 
is a consequence of a population-level 
distribution of the location of each cell's photoreceptor relative to the equatorial plane of the cell.

\textit{C. reinhardtii} strain CC125 was grown axenically in Tris Acetate Phosphate (TAP) 
medium, which provides it with the required nutrients. 
Cells were grown at $22$~$^\circ$C and synchronized in a light-dark cycle of $16/8\,$h 
($\sim\! 70\,\mu$E/m$^{2}$s) with constant shaking 
at $160$ rpm.  They were harvested in the exponential phase ($\sim\! 10^6$ cells/mL) 
when they are the healthiest and most motile. The suspension was typically diluted by a factor of
$20$ to avoid collective effects, placed 
in an open Petri dish (Falcon 353001, diameter $\sim\! 3.5\,$cm) and kept in a dark box for 10 minutes before conducting experiments to ensure 
that all cells start from the same condition. 

The experimental setup \cite{SM} consists of two collimated blue light beams 
($470$nm, ThorLabs COP1-A)
illuminating the Petri dish located at the center of the stage of an inverted microscope (Olympus, IX83), 
ensuring proper control of the light directions and avoiding light gradients over the imaging field 
(Fig. \ref{fig1}(a)). The light intensities were controlled by 
an LED Driver (Thorlabs DC4100) through their 
driving currents, and calibrated using a SpectraPen mini (Photon Systems Instruments).
We use a $4\times $ objective (field of view $3.7\times3.7 {\rm mm^2}$), and captured videos at $20$ fps using a digital camera (Hamamatsu 
Orca Fusion-BT C15440-20UP).  Image analysis used a 
combination of ImageJ and MATLAB to track the cells; trajectories were then linked and labeled 
employing the Crocker-Grier algorithm \cite{CrockerGrier}. 

Because the algal suspension is contained in a thin chamber 
and the lights illuminate the chamber at the very shallow
angle of $\sim\! 5^\circ$ with respect to the plane of the stage, the swimming is effectively two 
dimensional. As shown in Figs. \ref{fig1}(a,b),
the two beams are at angles $\pm\delta$ relative to the midplane, pointing toward the 
positive $x$-axis along the unit vectors
$\hat{\bf v}_\pm$, with $\eta=I_+/I_->1$ the intensity ratio of the two beams.  

{\it Chlamydomonas} cells, viewed from behind, spin counterclockwise around their 
posterior-anterior axis with
frequency $f_r=\vert\omega_3\vert/2\pi\sim 1.5-2\,$Hz \cite{SM}.  Because of shading by 
proteins behind
the photoreceptor, when a cell swims
at an angle $\varphi< -\delta$ (region I) or
$\varphi>\delta$ (region III) both lights illuminate the 
photoreceptor in the same half turn, but 
when $\vert\varphi\vert <\delta$ (region II), the 
photoreceptor is illuminated only by one light in each half turn.

Starting from a dark state in which cells swim randomly, highly directional swimming
occurs within $\sim\!\!\!\!~10\,$s of the start of illumination.  
The swimming paths appear as 
sinusoidal oscillations around linear motion because they are projections onto the $xy$-plane of
helices.  We obtain the swimming angle $\varphi_i$ for 
each of typically 200 - 600 paths from the best-fit straight line and 
report $\langle \varphi^*\rangle$ as
the ensemble average for each choice of $(\delta,\eta)$, with error bars representing the 
standard deviation of the fitted slopes.
Figure \ref{fig1}(c) shows that the data
are well-fit by the ``tangent law", where
the swimming direction is along the unit vector 
$\hat{\bf u}^*=(\cos\varphi^*,\sin\varphi^*)$ defined as an intensity-weighted average of the light vectors,
\begin{equation}
    \hat{\bf u}^*=\frac{\eta\hat{\bf v}_+ +\hat{\bf v}_-}{\vert\eta\hat{\bf v}_++\hat{\bf v}_- \vert} \ \ \ \ {\rm or} \ \ \ \ 
    \tan\varphi^*=\frac{\eta-1}{\eta+1}\tan\delta.
    \label{tangentlaw}
\end{equation}
For $\eta\gg 1$ (or $\eta\ll 1$) the trajectory aligns with the lower (or upper) 
light ($\varphi^*=\delta$ or $\varphi^*=-\delta$), while for equal light intensities ($\eta=1$) cells swim 
along the $x$-axis ($\varphi^*=0$).  This law appears to be valid for half-angles $\delta$ as large 
as $\sim 65^{\circ}-70^{\circ}$ (darkest blue inverted triangles in Fig.~\ref{fig1}(c)), although in this situation 
cells do not 
follow the average direction as accurately, as illustrated by the increasing size of the error bars as 
$\eta \rightarrow 1$ \cite{SM} . We hypothesize that such an intensity-weighted law should remain valid for more than 
two lights as long as the angle between the two furthest lights is small enough ($2\delta\lesssim 140^{\circ}$). 

\begin{figure*}[t]
\includegraphics[width=1.7\columnwidth]{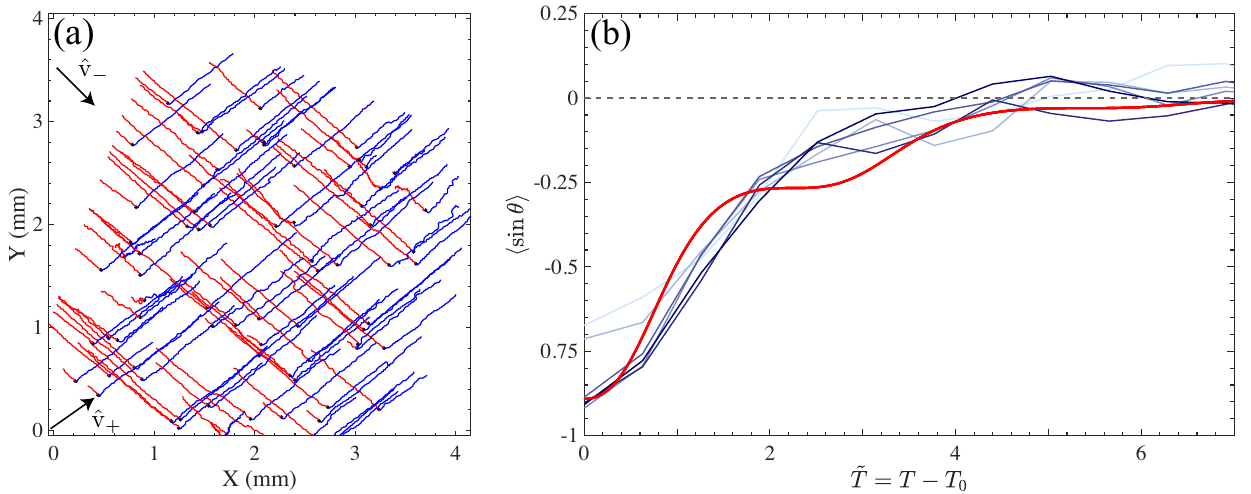}
\caption{Reorientation dynamics.  (a) Trajectories during a switch in light direction from $\hat{\bf v}_-$ (red) to 
$\hat{\bf v}_+$ (blue) for $\eta=1$.  Black circle indicates time of switch. 
(b) Reorientation angle $\theta$ during phototurn at intensities $I = 1.8, 3.7, 9.1, 17.7, 24.8, 33.9$ W\!/m$^{2}$,
color-coded from light to dark blue compared to
\eqref{fast} (red) averaged over initial phase of motion.}
\label{fig2}
\end{figure*}
% $(i_\pm = 5, 10, 25, 50, 75, 100)mA$, 

While the result \eqref{tangentlaw} makes no reference to biochemical processes in the cell, and is 
purely geometrical, we now show that it is a fixed point of a dynamical theory for 
\textit{Chlamydomonas} phototaxis \cite{Leptos2023}. This theory combines rigid-body dynamics and 
an adaptive model for the
angular rotation frequency $\omega_1$ around the body axis $\hat{\bf e}_1$ orthogonal to the flagellar beat plane  
due to asymmetries in beating of the two flagella in response to illumination of the photoreceptor.  
For a cell swimming in the
$xy$-plane with a photoreceptor along $\hat{\bf e}_2$ in the cell's equatorial plane
(Fig. \ref{fig4} in End Matter),
and with $T=\vert\omega_3\vert t$ a rescaled time, this dynamical system reduces to 
\begin{subequations}
\begin{align}
&\varphi_T=-P\sin T,\label{system1a}\\
&\textstyle P_{TT}+\frac{\alpha+\beta}{\alpha\beta}P_T+\frac{1}{\alpha\beta}P=\frac{1}{\beta}S_T.\label{system1b}
\end{align}
\label{system1}%
\end{subequations}
where the photoresponse variable $P=\omega_1/\vert\omega_3\vert$, and $\alpha$ and 
$\beta$ are the slow 
flagellar adaption time and fast response 
time made dimensionless with 
$\vert\omega_3\vert$, respectively, with $\alpha \gg \beta$ in experiments 
\cite{Leptos2023}. 
Under the assumption of additivity of light
stimuli, the phototactic signal $S= P^*\left[\eta J_+{\cal H}(J_+)+J_-{\cal H}(J_-)\right]$ is
given by the projections $J_\pm=-\hat{\bf o}\cdot 
\hat{\bf v}_\pm=\sin(\varphi\mp\delta)\sin T$ 
of the two lights on $\hat{\bf o}$, the outward normal to the eyespot, where $P^*=\omega_1^*/\vert\omega_3\vert$, with 
$\omega_1^*$ the 
peak turning rate around $\hat{\bf e}_1$. Here,  
the Heaviside functions ${\cal H}$ in $S$ represent
the effect of eyespot shading.  In the End Matter we show that averaging 
over the fast time scale of cellular spinning leads to the 
reorientation dynamics of the
swimming angle $\varphi$,
\begin{equation}
    \varphi_T=-\lambda
    \left[\eta\sin(\varphi-\delta)+\sin(\varphi+\delta)\right],
    \label{alignment_eom}
\end{equation}
where $\lambda\propto P^*$. While \eqref{alignment_eom} depends on cellular parameters via $\lambda$, 
its steady state solution $\varphi^*$ is the tangent law \eqref{tangentlaw}.

The dynamics \eqref{alignment_eom} has a Lyapunov function $V(\varphi)$ in the sense that $\lambda^{-1}d\varphi/dT=-dV/d\varphi$,
with \begin{equation}
    V(\varphi)=-\left[\eta\cos(\varphi-\delta)+\cos(\varphi+\delta)\right].
\end{equation}
A simple calculation shows that $V=-\hat{\bf e}_3\cdot (\eta\hat{\bf v}_++\hat{\bf v}_-)$, 
the  projection of the
total light vector onto the anterior pole of the cell, and that $V$ is a minimum at $\varphi^*$;
for negative phototaxis, 
a cell chooses a direction that minimizes the average light falling onto its anterior pole.

Moving on from the steady-state results in Fig. \ref{fig1}, we study how 
cells respond to a rapid switch in illumination between lights.  Figure \ref{fig2}(a) shows 
cellular trajectories for $2\delta= 70^{\circ}$.  These turns are quantified through the angle $\theta(T)$ between the local unit tangent 
$\hat{\bf t}$ to the trajectory
and the new light direction. Results for $5$ different light intensities are shown in Fig. \ref{fig2}(b),
which illustrates that complete reorientation occurs within just over one period of rotation ($T\simeq 2\pi$), with a
modest dependence on light intensity.
Within the adaptive theory, this rapid orientation corresponds to $P^*\sim 1$. 
When $\alpha\gg\beta$ and $\alpha\beta\sim 1$ as found 
in prior work \cite{Leptos2023}, the dominant balance 
in \eqref{system1b} gives $P\simeq S$ and thus $\theta_T=P^*\sin\theta \,{\cal H}(\sin T) \sin^2 \!T$.
In terms of the shift $\tilde{T}=T-T_0$ relative to the light-switching time $T_0$, we find
\begin{equation}
    \cos\theta(\tilde{T})=\tanh\left\{P^*\left[2\tilde{T}-\sin 2\tilde{T} \right]/4-C\right\}.
    \label{fast}
\end{equation}
where $C=\ln(\tan(\theta_0/2))$ and $\theta_0$ is the initial angle. To compare with the experimental data, 
we note that when the lights are switched the
cells are at random phases of their wiggly motion, and we average \eqref{fast} over a uniformly distributed
angle $\theta_0 \in [-110^\circ-\theta_i,-110^\circ+\theta_i]$, where $\theta_i=23^\circ$. The result
shown in Fig. \ref{fig2}(b) matches the data well, with $P^*\simeq 1.4$ as the sole fitting parameter.  

\begin{figure*}[t]
\includegraphics[width=2.0\columnwidth]{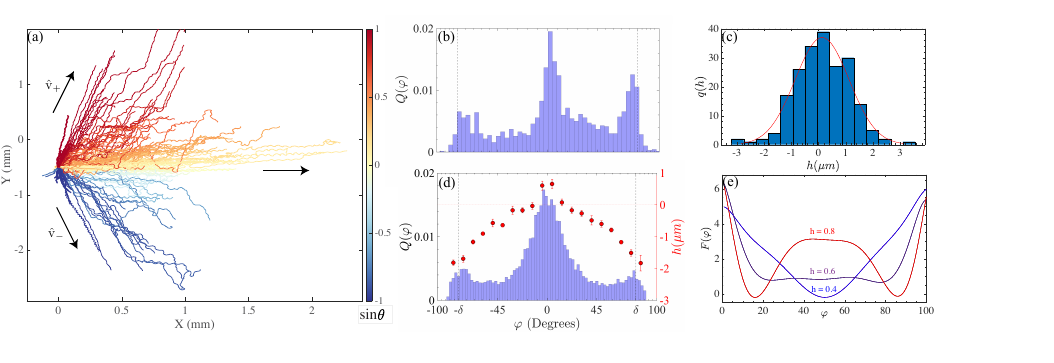}
\caption{Phototaxis at large angular light separation.  (a) Trajectories for 
$2\delta=162^\circ$ and $I_{\pm} = 3.7 {\rm W/m^{2}}$, showing negative phototaxis, 
following tangent law, and 
stochastic switching between directions.  Trajectories are color-coded by orientation of their end-to-end
vectors relative to the $x$ axis. (b) Probability distribution $Q(\varphi)$ of trajectory 
angles for $I_{\pm}= 3.7 {\rm W/m^{2}}$. (c) Distribution of 
eyespot offsets along with Gaussian fit. (d) Numerical $Q(\varphi)$ from model of stochastic phototaxis incorporating distribution of eyespot shifts from equator. 
(e) Effective free energy as a function of eyespot offset for $2\delta=162^\circ$. }
\label{fig3}
% {\color{cyan}RJ: What is the intensity of the lights in the experimental plots??}
\end{figure*}

This agreement validates the use of the
simplified model for 
$2\delta\simeq 180^\circ$ and $\eta=1$, where rapid reorientations are relevant.  
The three types of trajectories described in the Introduction, shown in 
Fig. \ref{fig3}(a) for $2\delta=162^\circ$, 
are quantified in Fig. \ref{fig3}(b) by partitioning each into $3$s segments ($\sim 5$ body rotations) and finding the average orientation angle $\varphi$ for each.  
The probability distribution $Q(\varphi)$ exhibits three peaks
corresponding to swimming away from either light or along the $x$-axis.

The theory discussed thus far 
cannot account for stable negatively phototactic swimming away from one light or the other when 
the photoreceptor is in the cell's
equatorial plane; the only stable fixed point is 
$\varphi^*=0$.  But if the photoreceptor is displaced from the equator 
by an angle $\gamma<0$, 
such motion is stable; due to 
photoreceptor shielding by the
eyespot, a cell swimming away from one light must 
rotate through $\gamma$ to sense the other.
Such rotations may arise from 
flagellar beating jitter \cite{Leptos2023}. 

Using a method based on light reflection from the eyespot \cite{Isogai2000,SM}, 
we measured the probability  distribution $q(h)$ of eyespot displacements $h$ from the midplane 
of $204$ cells, and 
show in Fig. \ref{fig3}(c) that it is well-fit by a Gaussian with mean $0.15\, \mu$m and standard 
deviation $1.0\, \mu$m (positive values correspond to eyespots closer to the flagella).  
A significant subpopulation has eyespots displaced
more than their own diameter from the midline.

It follows from the considerations above that
during the finite duration of an experiment,
cells with eyespots far below the midplane will 
have not had sufficient time to fluctuate
enough to see the other light, and thus will
swim away from one light or the other.  For 
those with eyespots far above the midline direct swimming away from either light is unstable.
And those with intermediate positions can exhibit stochastic hopping between the three choices.
To test the hypothesis that eyespot displacement is
the origin of the distribution in Fig. 
\ref{fig3}(b) we generalize the model by writing
a Langevin equation for the orientation with a shifted
photoreceptor. The dynamical system is
$\varphi_T=-P\sin T+\xi(T)$, where 
$\langle \xi(T)\rangle=0$, $\langle \xi(T)\xi(T')\rangle=2\tilde{D}_r\delta(T-T')$,
$\tilde{D}_r=D_r/\vert\omega_3\vert$ is the scaled rotational diffusion constant, the projections are
\begin{equation}
    J_\pm=\cos\gamma\sin(\varphi\mp\delta)\sin T-\sin\gamma\cos(\varphi\mp\delta),
\end{equation}
where $h=R\sin\gamma$, with $R=5\,\mu$m the cell
radius, and we use the limit $P\simeq S$ as in \eqref{fast}. In numerical studies we simulate
$N$ swimmers starting from random orientations in the plane, where each has an 
eyespot offset angle chosen from a Gaussian distribution
with parameters taken from experiment.  Figure 
\ref{fig3}(d) shows 
the resulting probability distribution function of trajectory angles, which 
strongly resembles the experimental one.  We find that 
the peaks are associated with different subpopulations 
of the cells.  Those with a large negative offset angle
$\gamma$ move directly away from one or the other light,
while those with a strongly positive offset follow the 
tangent law.  Cells with an offset close to the middle of these
two extremes stochastically switch
their trajectories.

By defining an effective free energy $F=-\ln Q$ from the measured
distributions in the numerical computations we find an underlying bifurcation in the decision-making 
process.  Simulations in which 
both $\delta$ and $h$ are fixed show (Fig. \ref{fig3}(e)) a transition from a single minimum in 
$F$ at small $h$ to a double-well
structure at larger $h$.  Thus, the observed three-peak distribution function in Fig. \ref{fig3}(d), 
obtained for an ensemble of cells with different eyespot offsets, 
reflects a superposition of 
one- and two-minimum free energies.
Further evidence for the existence of an underlying bifurcation at large angles is found in the 
increasing scale of fluctuations around the tangent law seen in Fig. \ref{fig1}(c) for large $\delta$ as 
$\eta\to 1$ \cite{SM}.

Since a continuously varying natural light field can be represented as a superposition of 
discrete sources, the results presented here suggest that the generalization of our model 
to one with $N\to\infty$ lights would imply that negatively phototactic cells would swim away from the brightest spot.  
This result provides a linkage 
between the line-of-sight and gradient-climbing approaches \cite{Williams2011,AndresBragado2018,Zhu2021}.
Issues for further study involve 
the possibility of a dynamic bifurcation \cite{LebovitzPesci1995} when
the lights are point sources, not at infinity, and 
the apparent position slowly varies as cells 
swim \cite{Sridhar2021},
as well as the effects of 
longer-term adaptations associated with photosynthesis
\cite{Moses2013,Arrieta2017}.

\begin{acknowledgements}
We thank Nelson Pesci for discussions. This work was supported in part by 
ANR JCJC funding (ANR-23-CE30-0009-01; RJ \& SR), 
the Lee and William Abramowitz Professorial Chair of Biophysics and a Royal Society Wolfson Visiting Fellowship (NSG), the John Templeton Foundation 
and the Complex Systems Fund at the University of Cambridge 
(AIP \& REG). 
\end{acknowledgements}

\textit{Data Availability-} The data that support the findings of this article are openly available \cite{Zenodo}.

\clearpage

\section{End Matter}
\label{endmatter}

\begin{figure}[b]
\includegraphics[width=0.55\columnwidth]{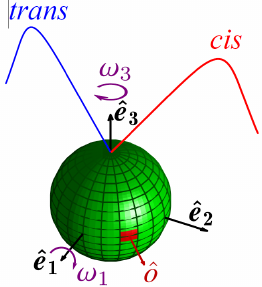}
\caption{Coordinate system of \textit{Chlamydomonas}.}
\label{fig4}
\end{figure}

The complete model introduced earlier \cite{Leptos2023} consists of the coupled dynamics of the
Euler angles $(\phi,\theta,\psi)$ describing rigid body motion and the adaptive dynamics 
of the angular rotational velocity $\omega_1$ about the axis $\hat{\bf e}_1$ that is orthogonal to the
plane of the flagellar beating.  In a system of units scaled by the magnitude of the spinning
angular velocity $\vert \omega_3\vert$ about the cellular axis $\hat{\bf e}_3$ that lies in the plane
of beating and is equidistant to the bases of the two flagella, the dynamics takes the form
\begin{subequations}
\begin{align}
\textstyle 
\phi_T &= \left(\hat{P}+P\right)\frac{\sin\psi}{\sin\theta}, \label{fd1} \\
\theta_T &= \left(\hat{P}+P\right)\cos\psi,\label{fd2} \\
\psi_T &= -1- \left(\hat{P}+P\right)\frac{\sin\psi\cos\theta}{\sin\theta},\label{fd3} \\
\beta P_T &= S-H-P, \label{fd4}\\
\alpha H_T &= S-H,\label{fd5}
\end{align}
\label{fulldynamicsscaled}%
\end{subequations}%
where $\hat{P}=\hat{\omega}_1/\omega_3$ and $P=\omega_1/\omega_3$ are the scaled intrinsic and 
photoresponse rotation frequencies
around $\hat{\bf e}_1$, respectively, $H$ is the hidden variable responsible
for adaptation and $\alpha$ and $\beta$ are the scaled response and adaptation times for the
photoresponse system.

In the absence of any light signal ($S=0$) the photoresponse variables vanish, we can describe analytically 
a weakly helical trajectory associated with a small $\hat{P}$ that deviates weakly from the $x-y$ plane.
Setting $\theta=\pi/2+\hat{P} \theta^{(1)}+\cdots$,
$\phi=\phi^{(0)}+\hat{P} \phi^{(1)} 
+\cdots$, and observe that $\psi=-T + 
{\cal O}(\hat{P}^2)$.  A simple calculation yields $\phi^{(1)}=\cos T$ and
$\theta^{(1)}=\sin T$, thus
\begin{subequations}
\begin{align}
x(T) &=U\left[T\sin\phi^{(0)}+
\hat{P} \cos\phi^{(0)}\sin T + \cdots\right]\\
y(T) &=-U\left[T\cos\phi^{(0)}-
\hat{P} \sin\phi^{(0)}\sin T + \cdots\right]\\
z(T) &= U\hat{P}\cos T + \cdots,
\end{align}
\label{helix}%
\end{subequations}%
the equation of a helix about a line 
in the $x-y$ plane oriented at angle 
$\phi^{(0)}$ with respect to the $y$-axis.
If we define $\hat{\bf u}=(\sin\phi^{(0)},-\cos\phi^{(0)})$ and 
$\hat{\bf u}^\intercal=(\cos\phi^{(0)},\sin\phi^{(0)})$, then the position of the swimmer is
\begin{equation}
    {\bf r}(T)=UT\hat{\bf u}+U\hat{P}\left[ \hat{\bf u}^\intercal\sin T +\hat{\bf e}_z\cos T \right],
\end{equation} 
making clear that the
helical radius is set by $\hat{P}$.
As this solution holds when $S=0$, we can view it
as the solution to the ``homogeneous" 
linearized version of 
Eqs. \eqref{fulldynamicsscaled}. 
 Hence,
the full solution of the 
linearized inhomogeneous problem is the sum of
\eqref{helix} and the particular solution
of the linearized inhomogeneous problem.  

In earlier work \cite{Leptos2023} we found two intrinsic time scales of relevance to 
a phototactic turn:
a short one associated the cell's spinning motion about the axis ${\bf \hat{e}}_3$ and a longer one 
for the reorientation of ${\bf \hat{e}}_3$ toward a collimated light beam.  In this section, we provide a systematic
derivation of the averaged dynamics based on the presumed smallness of the photoresponse variable $P$, focusing first on
the case of a single light source pointing along $-{\bf \hat{e}}_x$. 
To make analytical progress, we ignore the small helical component
of the motion, take the photoreceptor to be along $\hat{\bf e}_2$, and fix the Euler angle $\theta=\pi/2$ 
to neglect the small out-of-plane motion.  Finally, with $P$ small, we may approximate the third Euler angle by $\psi=-T$ 
since the time evolution of $\psi$ is dominated by rotations around 
$\hat{\bf e}_3$.  This yields the simplified model for the Euler angle $\phi$
\begin{subequations}
\begin{align}
\phi_T &= -P\sin T,\label{fulldyna} \\
\beta P_T &= S-H-P,\label{fulldynb} \\
\alpha H_T &= S-H, \label{fulldync}
\end{align}
\label{fulldynamics_new}%
\end{subequations}
We set $P^*$, the scaled maximum magnitude of the photoresponse, to be
\begin{equation}
    P^*=-\sigma\epsilon, \ \ \ \ {\rm with} \ \ \ \ \sigma=\pm 1, \ \ \ \ {\rm and}\ \ \ \ 0< \epsilon \ll 1,
\end{equation}
$\pm 1$ for positive/negative phototaxis.
Then the signal is 
\begin{equation}
    S=-\sigma\epsilon\cos\phi f(T) \ \ \ {\rm with} \ \ \ f(T)=\sin T\,\,{\cal H}\!\left(\sin T\right).
    \label{eq:simple1a}
\end{equation}
Noting that \eqref{fulldynb} and \eqref{fulldync} can be combined into a single second-order equation for $P$ \cite{Leptos2023},
\begin{equation}
\textstyle    {\cal L}P\equiv P_{TT}
    +\frac{\left(\alpha+\beta\right)}{\alpha\beta}P_T
    +\frac{1}{\alpha\beta}P=\frac{1}{\beta}S_T,
    \label{Peqn}
\end{equation}
we now seek a perturbative solution
to the dynamics \eqref{fulldyna} and \eqref{Peqn} in powers of $\epsilon$.

As the intrinsic rate of variation of $P$ given by $\alpha$ and $\beta$, is  ${\cal O}(1)$
it is natural to assume that when coupled to the slow variable $\phi$ the dynamical variables $\phi$ and $P$ 
can be written as functions of two variables, a fast one ($T$) and a slow one ($\tau=\epsilon T$).
Thus assuming the forms $\phi(T,\tau)$, etc. we have the rule
\begin{equation}
    \frac{d}{dT}\to \frac{\partial}{\partial T}+\epsilon\frac{\partial}{\partial \tau}.
\end{equation}
If we now propose perturbative expansions of the form
\begin{subequations}
\begin{align}
\phi(T,\tau)&=\phi^{(0)}(T,\tau)+\epsilon \phi^{(1)}(T,\tau)+\cdots, \\
P(T,\tau)&=P^{(0)}(T,\tau)+\epsilon P^{(1)}(T,\tau)+\cdots,
\end{align}
\label{pert1}%
\end{subequations}
then at ${\cal O}(\epsilon^0)$ we find
\begin{equation}
    P^{(0)}=0 \ \ \ \ {\rm and} \ \ \ \ \frac{\partial \phi^{(0)}(T,\tau)}{\partial T}=0,
\end{equation}
which implies that $\phi^{(0)}$ depends only on the slow variable $\tau$.
This solution corresponds to the fixed point of the underlying adaptive dynamics in the absence of a stimulus.

\begin{figure}[t]
\includegraphics[width=0.95\columnwidth]{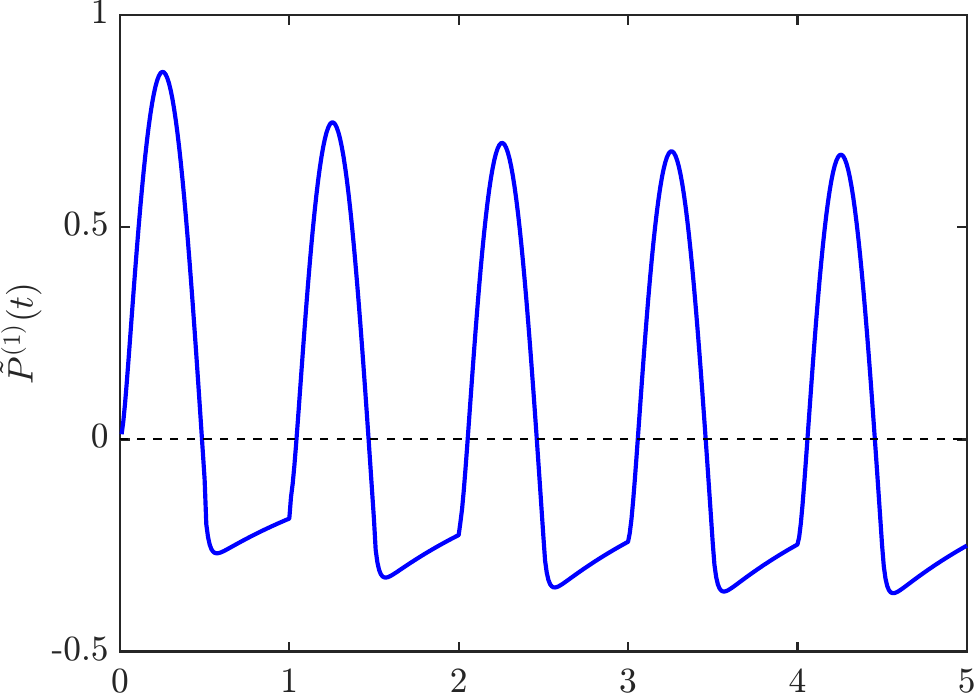}
\caption{The correction function 
$\tilde{P}^{(1)}$ for $\alpha=7$ and 
$\beta=0.14$.}
\label{fig5}
\end{figure}

At ${\cal O}(\epsilon^1)$ we find
\begin{subequations}
\begin{align}
\frac{\partial \phi^{(1)}}{\partial T}&=-\frac{\partial \phi^{(0)}}{\partial \tau}-\sin T \,P^{(1)},\label{eps1a} \\
{\cal L} P^{(1)}&=-\frac{\sigma}{\beta}\cos\left[\phi^{(0)}\left(\tau\right)\right] \frac{df(T)}{dT},\label{eps1b} 
\end{align}
\label{eps1}%
\end{subequations}
Note that the r.h.s. of \eqref{eps1a} does not depend on $\phi^{(1)}$ and thus functions as a forcing term.
Moreover, at this order the right-hand-side of \eqref{eps1b} is a functios of $T$, with $\tau$ playing the 
role of a parameter.  It is thus possible to solve \eqref{eps1b} independently of \eqref{eps1a}.
The l.h.s. of \eqref{Peqn} is simply a damped harmonic oscillator.  Since 
$[(\alpha+\beta)/2\alpha\beta]^2>1/\alpha\beta$, the Green's function is
\begin{equation}
    G(T,T')=\frac{\alpha\beta}{\alpha-\beta} {\cal H}(T-T')\left[e^{-(T-T')/\alpha}-e^{-(T-T')/\beta}\right].
\end{equation}
The term $df/dT$ on the r.h.s. of \eqref{eps1b} is $\cos T\,{\cal H}(\sin T)+\cos T\, [\sin T\,\delta(\sin T)]$,
whose second term makes no contribution to the convolution with $G$. 
With initial condition $P^{(1)}(0)=0$, the solution at a time $T$ such that $2n\pi < T \leqslant 2(n+1)\pi$ (i.e. 
$n-1$ full cycles have passed) is $P^{(1)}(T)=
-\sigma\cos\phi^{(0)}\tilde{P}^{(T)}(t)$, where
\begin{widetext}
\begin{equation}
\tilde{P}^{(1)}(T) =  
\begin{cases}
	\Lambda_1 B_n e^{-(T-2n\pi)/\beta} - \Lambda_2A_n e^{-(T-2n\pi)/\alpha}
	+ \Lambda_3\sin{T} + \Lambda_4\cos{T} ; & {\rm if} \ \ 2n\pi < T < (2n+1)\pi, \\
	\Lambda_1 B_{n+1} e^{-(T-(2n+1)\pi)/\beta} - \Lambda_2A_{n+1} e^{-(T-(2n+1)\pi)/\alpha}; 
 & {\rm if} \ \ (2n+1)\pi \leqslant T \leqslant 2(n+1)\pi,
\end{cases}
\label{tildePeqn}
\end{equation}
with $A_n=(1-r^{2n+1})/(1-r)$, $B_n=(1-q^{2n+1})/(1-q)$, $r=e^{-\pi/\alpha}$, $q=e^{-\pi/\beta}$, and
\begin{equation}
\textstyle
\Lambda_1= \frac{\alpha\beta}{(1+\beta^2)(\alpha-\beta)}, \ \ 
\Lambda_2= \frac{\alpha^2}{(1+\alpha^2)(\alpha-\beta)}, \ \
\Lambda_3= \frac{\alpha(\alpha+\beta)}{(1+\beta^2)(1+\alpha^2)}, \ \ 
\Lambda_4= \frac{\alpha(1-\alpha\beta)}{(1+\beta^2)(1+\alpha^2)}. 
\end{equation} 
\end{widetext}
For the realistic parameters $\alpha=7$ and
$\beta=0.14$ \cite{Leptos2023}, Fig. 
\ref{fig5} shows the function $\tilde{P}^{(1)}$ over several complete cycles.

Since we know that the physically relevant solution for the angle $\phi$ is bounded as 
$T\to\infty$, we must seek a bounded solution to \eqref{eps1a}, which can only be achieved
by removing any secular terms.  The freedom to do so is provided by the as yet unknown 
value of $\partial \phi^{(0)}/\partial\tau$.  The solvability condition is that the r.h.s. of \eqref{eps1a} be orthogonal to the nullspace of the l.h.s. operator.  As the nullspace is a constant, we
require only that the vanishing of the integral of the r.h.s. over any complete period from $T=2m\pi$ to $2(m+1)\pi$, for 
$m=0,1,\ldots$.  The necessary integrals were performed in earlier work \cite{Leptos2023}, from which
we deduce the equation of motion necessary for a bounded solution,
\begin{equation}
    \frac{\partial \phi^{(0)}}{\partial \tau}=\sigma\lambda\cos\phi^{(0)}.
    \label{phizero}
\end{equation}
where $\lambda=\Lambda_3/4$.
Expressed in terms of the angle $\varphi^{(0)}=
\phi^{(0)}-\pi/2$, with initial condition $\varphi_0=\varphi^{(0)}(0)$, we have
\begin{equation}
    \frac{\partial \varphi^{(0)}}{\partial \tau}=-\sigma\lambda\sin\varphi^{(0)},
    \label{phizero1}
\end{equation}
whose solution is
\begin{equation}
\varphi^{(0)}(\tau)=2\tan^{-1}\left[\tan\left(\frac{\varphi_0}{2}\right)e^{-\sigma\lambda \tau}\right].
\end{equation}
For positive phototaxis ($\sigma=+1)$, $\varphi\to 0$ as $\tau \to \infty$ (the cell
swims along $+\hat{\bf e}_x$, antiparallel to the light) whereas $\varphi \to \pi$ as 
$\tau\to\infty$ for negative phototaxis ($\sigma=-1$), as the cell swims parallel to the beam. 

The generalization of \eqref{phizero1} to the case of two lights at angles $\pm \delta$, leading to 
\eqref{alignment_eom}, is straightforward.

%\end{document}
\clearpage
\newpage
\setcounter{equation}{0}
\setcounter{figure}{0}
\setcounter{table}{0}
\setcounter{page}{1}
\makeatletter
\renewcommand{\theequation}{S\arabic{equation}}
\renewcommand{\thefigure}{S\arabic{figure}}
\renewcommand{\bibnumfmt}[1]{[S#1]}
\renewcommand{\citenumfont}[1]{S#1}
%%%%%%%%%% Prefix a "S" to all equations, figures, tables and reset the counter %%%%%%%%%%

\widetext

%\section{Supplemental Material}

%\renewcommand{\thesection}{S\arabic{section}}

%\renewcommand{\theequation}{S\arabic{equation}}

%\renewcommand{\thetable}{S\arabic{table}}

%\renewcommand{\thefigure}{S\arabic{figure}}

\section{Supplemental Material}

\section{S1. Fourier analysis of cell trajectories in 
single-light experiments}

\begin{figure}[b]
\centering
\includegraphics[width=1\textwidth]{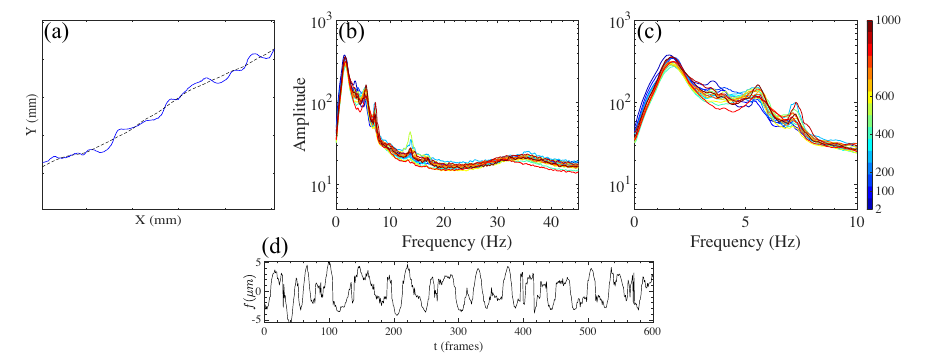}
\caption{Analysis of swimming trajectories. (a) Typical alga trajectory (solid blue)  with its mean computed 
from the SG Filter (dotted black). (b) Fourier spectra for different light intensities   (c)Enlargement of the first $10\,$Hz of the 
spectra shown in (c) (see colorbar). (d)Fluctuations of the trajectory shown 
in (a) around the mean trajectory computed from the SG 
filter.
}
\label{fig:S1}
\end{figure}

To understand in detail the effect of light 
intensity on the swimming of the cells, we first studied 
the trajectories in response to a single collimated light 
beam. Experiments were performed using a $\times 4$
objective at $90\,$fps in order to obtain well resolved
Fourier spectra of the motion. 
Because cell trajectories present modulations at frequencies below $0.5\,$Hz when cells follow a single light direction (Fig. 
\ref{fig:S1}(a), solid blue line), we 
first applied a Savitzy-Golay (SG) filter to the tracks in 
order to define properly the underyling mean motion around which are fluctuations associated with the 
the helical motion of the cells. The parameters of 
the SG filter were fixed to window size of $49$ frames and polynomial order as $1$. The 
SG filter then yields a mean trajectory (dashed black line in Fig. \ref{fig:S1}(a)) around which 
we can analyze the fluctuations of the filtered path, as shown in Fig. 
\ref{fig:S1}(b). The Fourier transforms for several light intensities shown in Figs. \ref{fig:S1}(c,d) 
are obtained by first trimming all trajectories at a given intensity to a 
common duration of $1300$ frames, computing the FFTs of each, and then averaging those spectra. 
We observe a clear peak at $1.5 - 2\,$Hz, common to all 
intensities, that corresponds to the frequency of helical 
swimming.  Interestingly, the structure of the spectra above $2\,$Hz appears to depend on the light intensity; 
at low intensity there is a small peak (or ``shoulder") between $3.5$ and $4\,$Hz that disappears as the light intensity is 
increased, while other peaks appear at just below $6$ and $8\,$Hz. 
Other peaks are also present at larger frequencies (between $10$ and $20\,$Hz) as shown in 
Fig. \ref{fig:S1}(c). The biological explanation 
for these high-frequency peaks and their dependence on light
intensity remains unclear. The broad peaks seen between $30$ and $40\,$Hz, corresponding to the small, fast 
fluctuations visible in Fig. \ref{fig:S1}(d),
arise from detection noise of the positions of cells.

\section{S2. Tangent Law}
\subsection{A. Current-to-Power conversion}

\begin{figure}[h]
    \centering
    \includegraphics[width=0.5\linewidth]{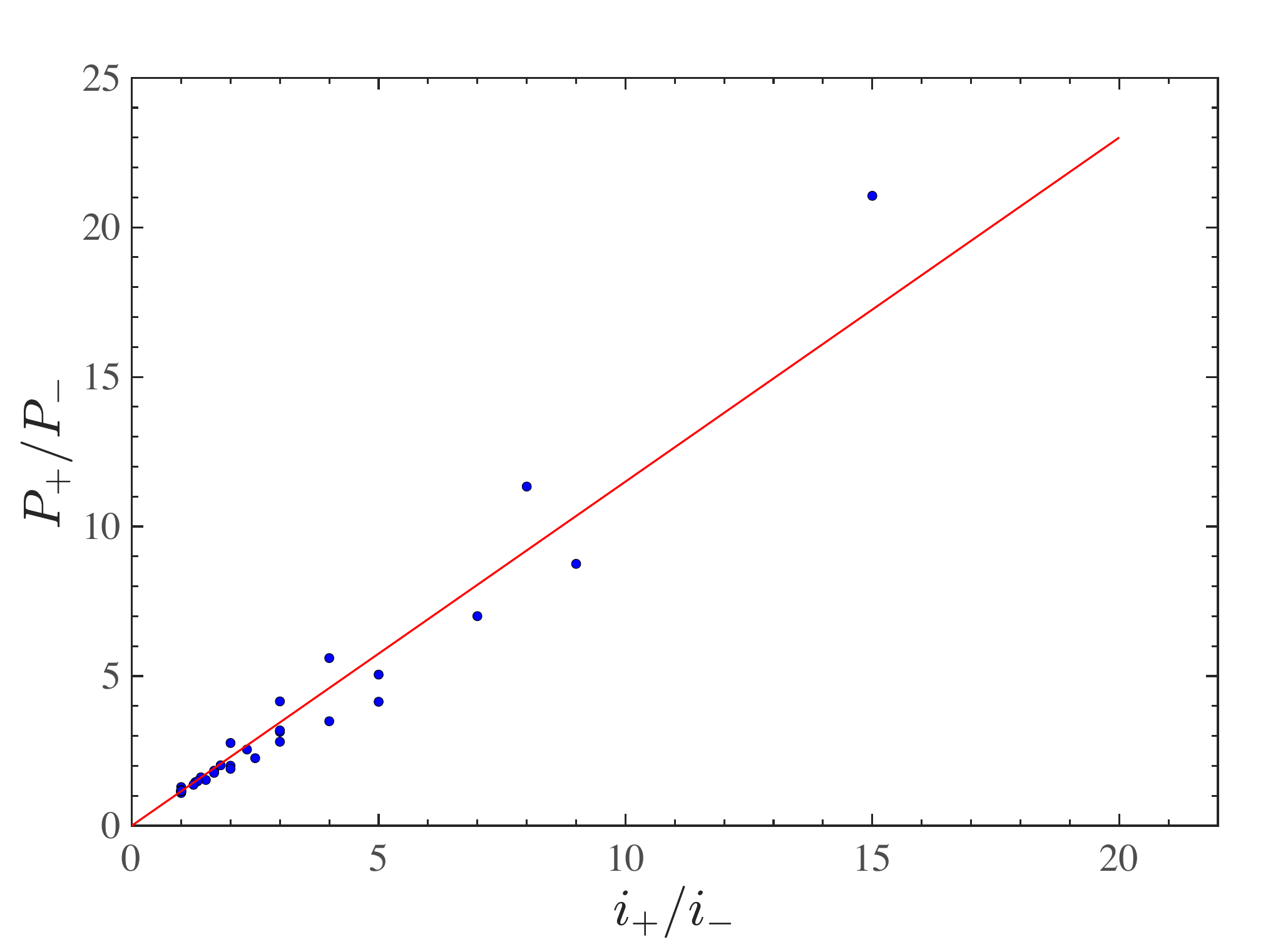}
    \caption{Calibration curve for $2\delta = 49.5^{\circ}$ with the linear fit shown in red 
    (here $P_+/P_-=1.15\, i_+/i_-$).}
    \label{fig:S2}
\end{figure}

The LED driver controls the electrical current sent to the device. To
calibrate the power transmitted by the LED, we used a power 
meter (Thorlabs controller PM100D with S130VC power sensor) that is held perpendicular to the 
collimated light beam, at the center of the stage, where we image the 
algae. We perform a linear fit to the plot of power versus 
current, the slope of which yields the calibration 
coefficient that is used in converting the ratio $i_+/i_-$
of the electrical currents to a ratio of light power or light intensity $\eta=P_+/P_-=I_+/I_-$.
This measurement is repeated over all experiments with
different angles.  An example of a calibration curve is shown in Fig. \ref{fig:S2}.

For information, with our blue LED sources the conversion of light intensity in ${\rm W/m^2}$ to ${\rm \mu E/m^2/s}$ is given by: $ 1 {\rm W/m^2}=3.96 {\rm \mu E/m^2/s}$.

\subsection{B. Experimental determination of the angles for the tangent law}

\begin{figure}[h]
    \centering
    \includegraphics[width=\linewidth]{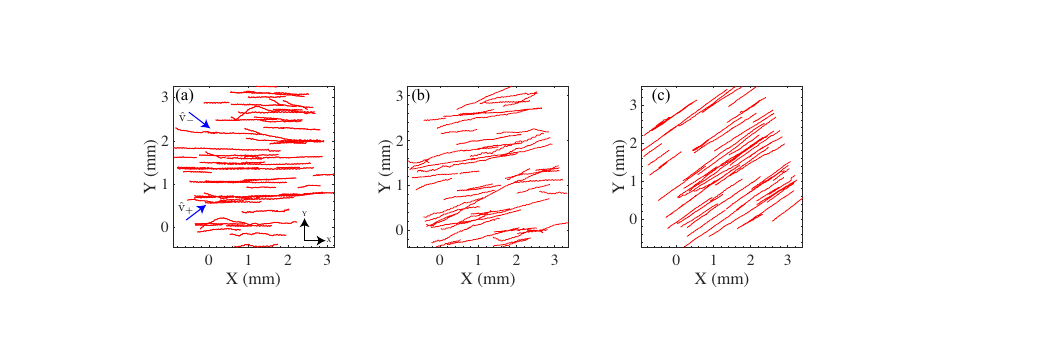}
    \caption{Trajectories for $2\delta=76.8^{\circ}$ under different light intensity ratios: (a) $\eta = 0.74$, (b) $\eta = 2.9$, 
    (c) $\eta = 16.6$. }
    \label{fig:S3}
\end{figure}

To obtain the tangent law, two series of experiments were performed for each fixed position of the two lights. 
We first performed tracking experiments with each light turned on independently, which allows us to get an 
accurate value of the angle $2\delta$ between the lights in the frame of the camera. This relies on the fact 
that cells do follow accurately the light propagation direction when a single light source is on. The 
ensemble average of the swimming direction of many 
cells, each obtained by a linear fit to its trajectory, is then used to define the light propagation 
directions. We then performed experiments with the two lights turned on to extract the ensemble average angle 
of swimming $\langle \varphi^*\rangle$ as a function of $\eta$, obtained also by a linear fit to the 
trajectories. Figures \ref{fig:S3}(a-c) show representative trajectories of the cells for a fixed $2\delta=76.8^{\circ}$ 
at different values of $\eta$, illustrating how the  
swimming angle $\varphi^{*}$ changes as $\eta$ increases.

\subsection{C. Analysis of the standard deviation of $\varphi^*$}

\begin{figure}[h]
\centering
\includegraphics[width=0.6\linewidth]{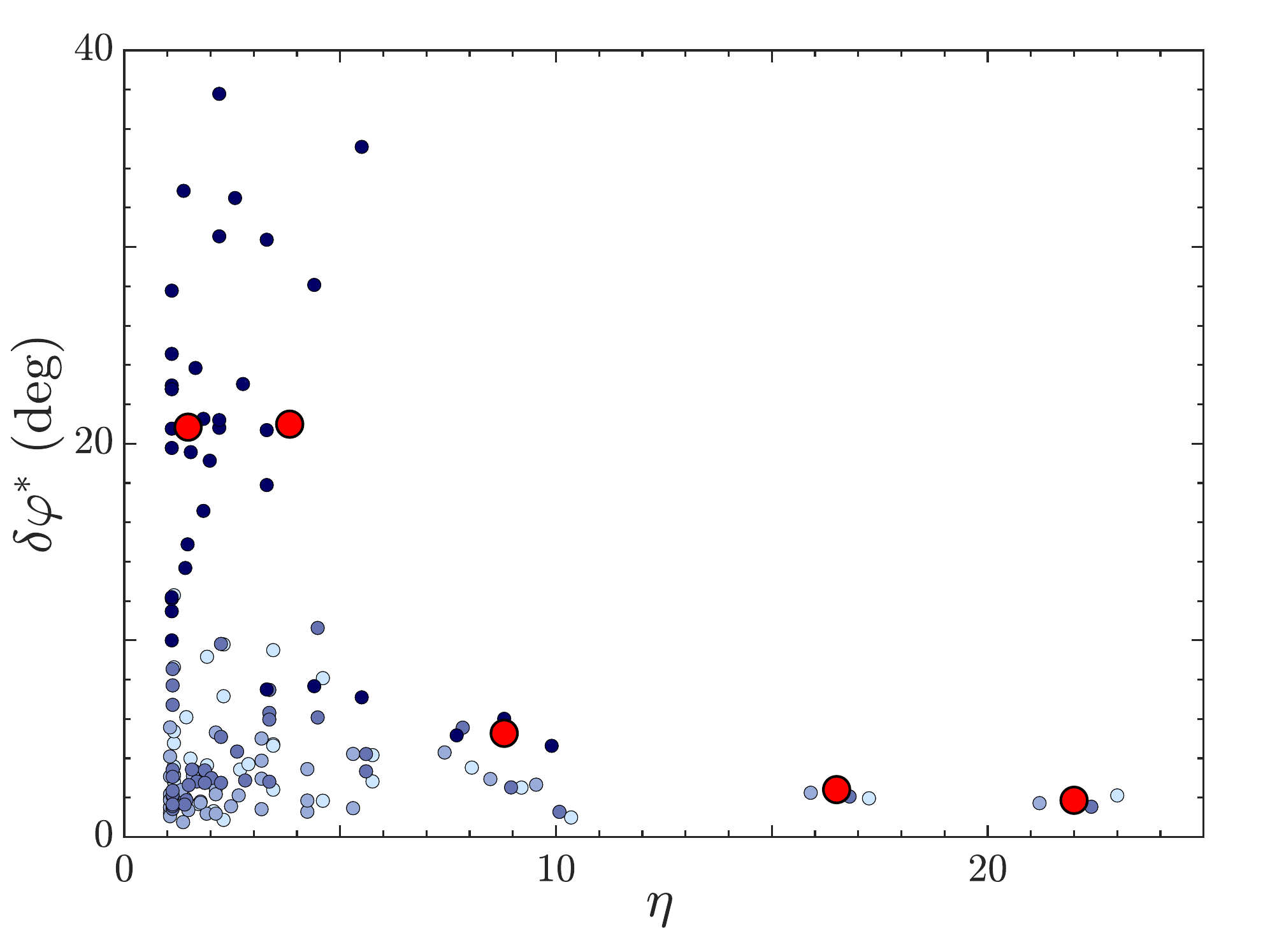}
\caption{Standard deviation of swimming direction $\delta \varphi^*$ among the cell population 
for all $\delta$ (color-coded from light to dark blue for increasing $\delta$, as Fig. 1(c) of main text). 
For the largest angle $2\delta\approx 135^{\circ}$ (dark blue) the standard deviation increases drastically 
as $\eta\to 1$ (for $\eta \lesssim 5$). This is shown by the red circles, obtained by binning the data with 
similar values of $\eta$.}
    \label{fig:S4}
\end{figure}

As the angle between the two lights is increased to $2\delta\approx 135^{\circ}$, we observed 
that the fluctuations in cell swimming direction drastically increased as $\eta$ gets closer to 1 (Fig. \ref{fig:S4}). 
Although the tangent law can here still be considered as valid on average, the ability of cells to follow the 
intensity-weighted average light direction is impeded at such large angles. This is a first sign that a 
bifurcation will be observed for larger angles, as we see for $2\delta=162^{\circ}$. 
As discussed in the main text, the three-peak distribution of trajectory angles arises from a superposition of the 
properties of subpopulations of cells.  For this reason, there should be no sharp transition or true bifurcation as $\delta$ 
is increased. 

\section{S3. Light-switching experiments}

\subsection{A. Obtaining the switch time $t_0$}

\begin{figure}[h]
    \centering
    \includegraphics[width=0.5\linewidth]{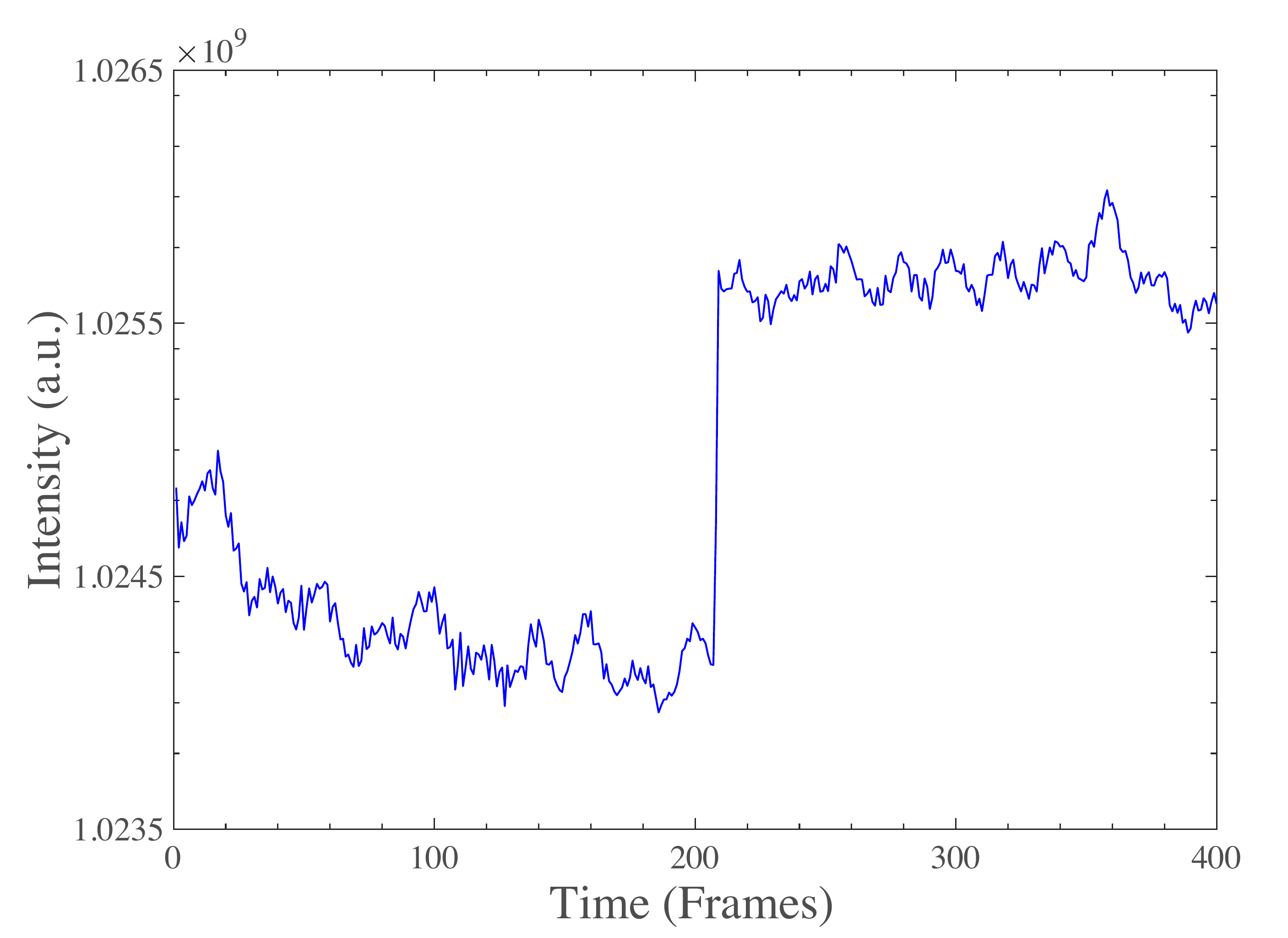}
    \caption{Total frame intensity over 
    the duration of an experiment for $i_+=i_- = 50\,$mA.}
    \label{fig:S5}
\end{figure}

The experiments were performed at a frame rate of $20$ fps, using the same $4\times$ objective as for experiments 
above.  Movies were recorded for a period of $20$ seconds, within which the light was switched manually 
at $\sim 10$s after 
the start of the recording. 
To determine the video frame corresponding to the switching of the light, and hence the switch time $t_0$, 
we located the jump in total intensity 
of the video frames that arises from the very small intensity difference of the two light sources.
Fig. \ref{fig:S5} shows a typical intensity trace with such a jump.

\subsection{B. Dispersion of initial value of $\sin\theta$ in light-switching experiments}

\begin{figure}[h]
    \centering
    \includegraphics[width=0.75\linewidth]{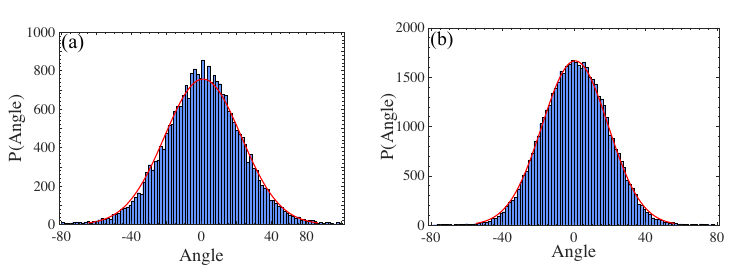}
    \caption{Distribution of swimming directions prior to the switch time $t_0$ in light-switching experiments. The distributions are centered around the direction of light propagation $\hat{\bf v}_+$. (a) A Gaussian fit 
    (red curve) at low intensity  $I_+ = 1.8 {\rm W/m^{2}}$ ($i_+=5$mA) yields the average angle $\mu = 0.78^{\circ}$ and standard deviation $\sigma = 21.89^{\circ}$. (b) At the larger intensity $I_+ = 63.1 {\rm W/m^{2}}$ ($i_+=100$mA), $\mu = 0.63^{\circ}$ and $\sigma = 18.52^{\circ}$.}
    \label{fig:S6}
\end{figure}

In the light-switching experiments, the angle of swimming of the cells at the switch time 
$t_0$ does not correspond exactly to the light propagation direction $\hat{\bf v}_+$. This is a consequence 
of the helical motion of the cells that makes their orientation oscillate around the average direction 
$\hat{\bf v}_+$. Typical distributions of orientations before the switch are shown in Fig. \ref{fig:S6} for 
$I_+ = 1.8 {\rm W/m^{2}}$ ($i_+=5$mA) (panel a) and $I_+ = 63.1 {\rm W/m^{2}}$ ($i_+=100$mA) (panel b). In order to compare with the theory (Fig. 2b and Eq. (5) 
of the main text), we need to take into account this variation in orientation at $t_0$, which makes the 
average value $\langle \sin \theta(T_0)\rangle$ larger than $-\sin 2\delta\approx -0.94$. The width of these 
distributions decreases as the light intensity increases (Fig. \ref{fig:S6}), consistent with the ordering of the 
curves in Fig. 2(b) of the main text at the switch time $t_0$.

\section{S4. Bifurcation}

\subsection{A. Calibrating light intensity}

For the bifurcation experiments at large angles $2\delta$ between the lights, we require the intensity of 
each light to be exactly the same. Otherwise, there would be a strong asymmetry in the distribution of orientations $Q(\varphi)$ because cells are highly sensitive to even minute differences between the two light intensities. 

To ensure that $\eta$ is as close to $1$ as possible, we calibrated the two LEDs using a 
Spectra Pen Mini (Photon System Instruments). A custom stage was built to hold the sensor at its center, exactly 
where the cells are imaged in the Petri dish. We then adjusted the electrical current sent to one of the lights 
to match exactly the intensity of the second light. This protocol leads to distributions $Q(\varphi)$ that are quite symmetric, although a slight asymmetry persists.  

\subsection{B. Calculating the distribution of swimming directions $Q(\varphi)$}

To obtain a detailed quantification of the swimming direction 
of the cells, we divide each 
trajectory into segments consisting of $5$ full rotation 
periods ($60$ frames $=3$s). From these segments we calculate 
the local angle ($\varphi$) with respect to the average direction ($\hat{\bf v}_+ + \hat{\bf v}_-$),
yielding the distribution of orientations ($Q(\varphi)$, Fig. 3(b) in the main text). The same analysis was performed on the simulated trajectories (Fig. 3(d) in the main text).

\subsection{C. Eyespot imaging}
\begin{figure}[h]
\centering
\includegraphics[width=0.8\linewidth]{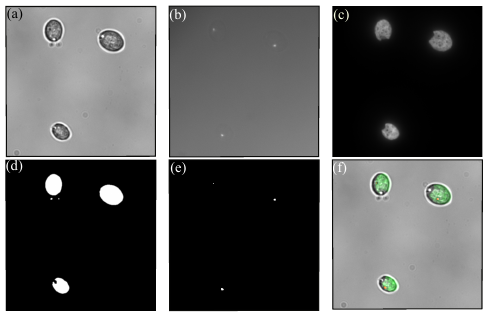}
\caption{Determining the eyespot location. 
(a) Brightfield image of the cell bodies, (b) eyespots (bright spots) in reflection microscopy, 
(c) chloroplast visualized using epi-fluorescence microscopy of the chlorophyll molecules, 
(d) binarized image of panel a, (e) binarized image of panel b, 
and (f) merged image of panels a, e and c in gray, red and green color respectively.}
    \label{fig:S7}
\end{figure}

In order to determine the distribution of eyespot 
locations in a population of \textit{C. reinhardtii}, we first immobilized the cells on a microscope slide coated with 
polylysine (Electron Microscopy Sciences, \#63410-01) by placing a $12\,\mu$L droplet onto the slide and covering it 
with a coverslip. We then combined the information obtained from three different types of imaging with a $40\times$ 
objective (Olympus, UPlanSApo, 40$\times$/0.95): brightfield microscopy, reflection microscopy, and epi-fluorescence 
microscopy (methods detailed below). Brightfield imaging (Fig. \ref{fig:S7}(a)) was used to characterize the cell shape. 
Reflection microscopy with an orange (580nm) light (Fig. \ref{fig:S7}(b)) was used to image the eyespot as a bright spot 
making use of the highly reflective properties of the carotenoid layer of the eyespot at this wavelength. Epi-fluorescence 
microscopy was finally used to image the chloroplast and obtain the polarity of the cells. Acquiring about 
20 images with these methods, we obtained statistics of the eyespot location of $\sim 200$ cells. 
A superposition of the three types of images is shown Fig. \ref{fig:S7}(f) with the brightfield image in grey, the 
eyespot in red and the chloroplast in green. 

The brightfield images were binarized, followed by dilation and erosion to obtain 
ellipsoidal cell bodies (Fig. \ref{fig:S7}(d)) from which we extracted the centroid, the major and minor axis, 
the cell area (to filter out cells too close to each other to be detected as single cells) and the bounding 
rectangle of the features. The images from reflection microscopy were also simply binarized (Fig. \ref{fig:S7}(e)) 
to extract the eyespot position. From the information of the eyespot position and the major axis of the cells we 
obtained the distance of the eyespot from the cell equator, without knowledge of the direction (i.e. closer to the flagella 
or further away). This missing information was obtained from the epi-fluorescence images of the chloroplast which makes 
a U-shape at the back of the cells (Fig. \ref{fig:S7}(c)). Extracting the center of mass of the fluorescence signal 
allowed us to know the polarity of the cells. To do so we applied the binarized brightfield images as masks to the 
epi-fluorescence images, then cropped around each cell (using the bounding rectangle) and finally computed the center of 
mass of the fluorescence signal.

\textbf{Details of the imaging methods.}
All images were acquired on an Olympus IX83 inverted microscope equipped with a Hamamatsu Orca Fusion-BT camera (C15440-20UP). 

\textit{Brightfield.}
Brightfield images were captured using a red filter at $624\,$nm (BrightLine, FF01-624/40-25) to ensure cells remained still by avoiding phototactic responses. 

\textit{Reflection microscopy.}
White light from a Lumencor SOLA-VN LED source was sent through the back of the microscope into a filter cube containing an excitation filter at 580nm (BrightLine, FF01-580/60-25) and a glass slide acting as a semi-reflective surface to redirect light towards the sample and let the reflected light from the eyespots to reach the camera. 

\textit{Epi-fluorescence microscopy.}
White light from a Lumencor SOLA-VN LED source was sent through the back of the microscope into a filter cube containing an excitation filter at 447nm (BrightLine, FF02-447/60-25), a dichroic mirror at 605nm (Thorlabs, DMLP605R) and an emission filter at 697nm (BrightLine, FF01-697/58-25) to collect the fluorescence signal of the chlorophyll molecules in the chloroplast.

\section{S5. Captions for Supplementary Movies}

\textbf{Supplementary Movie 1:} Cells following the tangent law, at various values of $\eta = 1.2, 2.9, 16.6$. Video captured at 4x magnification, 20fps, played in real time. \\

\textbf{Supplementary Movie 2:} On changing the light, we see the fast turns that the cells make when the light is switched from $\hat{\bf v}_{+}$ to $\hat{\bf v}_{-}$. Video captured at 4x magnification, 20fps, played in real time. \\

\textbf{Supplementary Movie 3:} At bigger angle of $2\delta = 162^\circ$ and $\eta=1$ different subpopulations of cells emerge: cells following either light, cells following the tangent law and cells switching stochastically between these two. Video captured at 4x magnification, 20fps, played in real time.

\end{document}